\documentclass[10pt,a4paper]{article}
\usepackage{graphicx}
\usepackage{tabularx}
\usepackage{multicol}
\usepackage{tabto}
\usepackage{url}
\usepackage{enumitem}
\usepackage{textcomp}
\usepackage{eurosym}
\usepackage{wasysym}
\usepackage[utf8]{inputenc}
\usepackage{flowchart}
\usepackage[labelfont=bf]{caption}
\usepackage[yyyymmdd]{datetime}

\newcolumntype{L}[1]{>{\raggedright\arraybackslash}p{#1}}
\newcolumntype{C}[1]{>{\centering\arraybackslash}p{#1}}
\newcolumntype{R}[1]{>{\raggedleft\arraybackslash}p{#1}}

\usetikzlibrary{shapes,arrows.meta,chains,external}

\newcommand{\eat}[1]{}

\newcommand{\ts}{
    \tikzset{>={Latex[width=3mm,length=3mm]}}
    \tikzstyle{line} = [draw, ->, >=latex, ultra thick]
    \tikzstyle{circ} = [
        circle,
        align=center,
        text width=2em,
        text centered,
        inner sep=1mm,
        outer sep=1mm,
        minimum width=0cm,
        minimum height=0cm
    ]
    \tikzstyle{rect} = [
        rectangle,
        align=center,
        text width=2em,
        text centered,
        inner sep=1mm,
        outer sep=1mm,
        minimum width=0cm,
        minimum height=0cm
    ]
    \tikzstyle{noshape} = [text width=5em, text centered, minimum height=5em]
}

\begin{document}

\bibliographystyle{plain}
\thispagestyle{empty}

\NumTabs{6}

\begin{center}
\large{\bf Can Cryptocurrencies Preserve Privacy and Comply with Regulations?}\\
\end{center}
\begin{center}
\begin{minipage}[t][][t]{0.44\linewidth}
\begin{center}
\large{\bf Geoff Goodell}\\
Centre for Blockchain Technologies\\
University College London\\
\texttt{g.goodell@ucl.ac.uk}
\end{center}
\end{minipage}
\begin{minipage}[t][][t]{0.44\linewidth}
\begin{center}
\large{\bf Tomaso Aste}\\
Centre for Blockchain Technologies\\
University College London\\
\texttt{t.aste@ucl.ac.uk}
\end{center}
\end{minipage}
\end{center}

\begin{abstract}

Cryptocurrencies offer an alternative to traditional methods of electronic
value exchange, promising anonymous, cash-like electronic transfers, but in
practice they fall short for several key reasons.  We consider the false choice
between total surveillance, as represented by banking as currently implemented
by institutions, and impenetrable lawlessness, as represented by
privacy-enhancing cryptocurrencies as currently deployed.  We identify a range
of alternatives between those two extremes, and we consider two potential
compromise approaches that offer both the auditability required for regulators
and the anonymity required for users.

\end{abstract}

\section{Introduction}
\label{s:intro}

The surveillance economy has arrived~\cite{zuboff2015}.  The popularity of
online service platforms has enabled service providers to collect, aggregate,
and analyse data about the behaviour of individuals with a volume and scope
never before possible.  Data brokers have created a marketplace for exchanging
information about individuals that can be used to link their various online
actions, including but not limited to financial transactions.  Such
information, including the reuse of credentials over successive transactions,
can be used to link the transactions to the transacting
parties~\cite{rieke2016,beckett2014}.  Such a linkage can greatly simplify
successive transactions, reducing costs for the provider and improving customer
experience.  However, the potential for monitoring profoundly influences the
everyday behaviour of individuals as they conduct their various
activities~\cite{schep2017}.  The value of such control is reflected in an
emerging marketplace for record linkage via \textit{entity resolution}, which
seeks to determine the specific individual person associated with any given
activity and, correspondingly, the history of activities associated with any
given individual person~\cite{waldman,pitneybowes}.

In the context of longstanding arguments that privacy is a public
good~\cite{warren1890,alrc2014,fairfield2015,gaur2017}, it is worthwhile to
consider whether such practices may serve to exacerbate social inequity by
restricting the ability to transact privately to those with sufficient wealth
and power~\cite{cole2014,hess2017}.  Financial transactions are no exception,
since they reveal information about not only the volume and recipients of
individuals' purchases and remittances but also their patterns, location
histories, social networks, and so on.  Modern retail banking creates a kind of
panopticon for consumer behaviour, ultimately promising to implement a
mechanism that binds all of the financial activities undertaken by an
individual to a single, unitary identity.  Consumers have legitimate reasons to
resist such surveillance, particularly in cases wherein monitoring is carried
out without their knowledge and judgments based upon such monitoring are used
to disincentivise or punish legitimate activities.  The risk to consumers
increases with the ever-increasing share of financial transactions that are
performed electronically.  The increasing capability of third parties to
aggregate and analyse data about retail financial transactions fundamentally
changes the relationship between individuals and their financial institutions.

Cryptocurrencies seem like a natural alternative for exchanging value that can
avoid the watchful eye of state actors, powerful corporations, hackers, and
others who might be well-positioned to build a dossier of one's activities.
However, a lack of appropriate regulation generally burdens cryptocurrency
users with practical limitations and risks.  The risks include the lack of
financial products and services, the inability to earn interest, basic consumer
protection, and the absence of legal infrastructure for adjudicating disputes.
China has also restricted the use of cryptocurrency exchanges as a means of
addressing capital outflows~\cite{chen2018}.

Additionally, most cryptocurrencies are not as privacy-enhancing as is commonly
perceived, and as state actors attempt to erect a cordon around criminal
activity that relies upon cryptocurrencies, some governments have actively
sought to undermine the adoption of cryptocurrencies that are most respectful
of an individual's privacy.  For example, the Financial Services Agency in
Japan pressured cryptocurrency exchanges to drop privacy-enhancing tokens such
as Monero~\cite{adelstein2018}, one of the largest cryptocurrency exchanges in
South Korea subsequently delisted privacy-enhancing tokens~\cite{dixit2018},
the US Department of Homeland Security specifically called for methods to
circumvent privacy protections in privacy-enhancing
cryptocurrencies~\cite{dhs2018}, and the UK Financial Conduct Authority offered
guidance indicating its intention to ``prevent [rather than simply prosecute]
the use of cryptoassets for illicit activity''~\cite{cryptoassets-taskforce}.
Furthermore, whether or not these government initiatives succeed in preventing
untraceable cryptocurrencies from achieving mainstream adoption, even the most
private cryptocurrencies suffer from the arguably intractable governance
challenges associated with building a decentralised network that respects the
interests of its users.

Following the G20 summit held in Buenos Aires in 2018, leaders resolved to
``regulate crypto-assets for anti-money laundering and countering the financing
of terrorism in line with FATF standards''~\cite{g20-2018}.  Earlier, the UK
Parliament had published a report citing lack of consumer protection and
regulated marketplaces for crypto-assets as major drawbacks associated with
cryptocurrencies as a medium of exchange, also noting the possibility that
anonymous transactions might promote money laundering is a significant
perceived risk, despite that the UK National Crime Agency had assessed the risk
as low~\cite{parliament2018}.  So we have reached an impasse, with institutions
demanding control and countenancing surveillance at one extreme, and
cyberlibertarians demanding privacy at the expense of regulation on the other.

\begin{table}
\begin{center}
\sf\begin{tabular}{|L{4cm}|}\hline
Robust to cyberattacks      \\
Usable without registration \\
Unlinkable transactions     \\
Electronic transactions     \\
Suitable for taxation       \\
Can block some illicit uses \\
Can be denominated in       \\
\vspace{-4.5mm}\hspace{3mm} units of fiat currency \\
\hline\end{tabular}\rm

\caption{\textit{Desiderata for an electronic payment method.}}

\label{t:desiderata}
\end{center}
\end{table}

Let us commence this article by introducing a set of ``desiderata'': properties
that a payment system should have.  They are listed in
Table~\ref{t:desiderata}.  In addition to usefulness, security, and privacy, we
also note the various arguments for money as a ``valid payment for all
debts''~\cite{royalmint}, as a means of funding government activity through
taxation~\cite{forstater2004}.  We discuss how we can reframe our requirements
such that we might achieve a parsimonious set of regulatory objectives while
also respecting privacy and fulfilling the desiderata.  We follow and extend
the ongoing discussion of how to regulate cryptocurrency
payments~\cite{hughes2015,tasca2018}, with a view toward respecting human
rights~\cite{goodell2018}.

The paper is organised as follows.  In the remainder of the first section, we
discuss the regulatory context surrounding modern retail financial
transactions, and we introduce cryptocurrencies as a prospective substitute for
regulated payments.  In the second section, we compare and contrast three
methods of conducting financial transactions online: modern, regulated retail
banking; classic cryptocurrencies such as Bitcoin; and privacy-enabling
cryptocurrencies such as Monero.  In the third section, we introduce two
candidate approaches that each offer individuals a verifiable means of
transacting privately and also provide suitable mechanisms by which
institutions can enforce regulatory compliance.  In the final section we
conclude with a discussion of the opportunities and tradeoffs.

\subsection{Institutional Posture}

Banks and other financial intermediaries in many jurisdictions around the world
are subject to anti-money laundering (AML) or ``know your customer'' (KYC)
regulations that require them to collect data on individual accountholders and
others who make use of their services~\cite{govuk2014,amlkyc}.  The penalties
for non-compliance are potentially severe.  In recent years, banks have
dedicated significant resources to building and maintaining compliance
infrastructure, evidenced by the thousands of employees that they have hired to
monitor ``high-risk'' transactions, as well as ``tens of thousands of costly
customer calls every month to refresh KYC documents''~\cite{breslow2017}.

An international organisation named the Financial Action Task Force (FATF) was
established by the G7 in 1989 as a trans-national effort to monitor financial
activities, with the stated purpose of investigating and preventing money
laundering and terrorist financing~\cite{fatf-recommendations}.  FATF provides
one of the mechanisms by which AML/KYC regulations in different jurisdictions
are promulgated and coordinated.  FATF also publishes a blacklist of nations
who fail to enforce rules that facilitate the identification and investigation
of individual accountholders, with the purpose of coordinating sanctions that
force blacklisted nations to conform~\cite{fatf2013}.

The financial regulations imposed by economically powerful jurisdictions such
as the United States and the European Union share common features.  In the US,
AML regulations provide for customer identification and monitoring, as well as
the reporting of suspicious activities~\cite{sec2017}.  In the EU, Directive
(EU) 2018/843 (``5AMLD'') requires that every financial transaction must be
associated with an account, and that every account must be associated with a
strongly identified responsible individual~\cite{eu5amld}.  The directive also
significantly reduces the maximum allowed value for prepaid cards and
stipulates that remote transactions above EUR 50 must be accompanied by
customer identification~\cite{eu5amld}.  Note that 5AMLD specifically includes
cryptocurrencies as subject to its prescribed regulations on financial
transfers.

Although the systematic collection of identifying information for individual
accountholders might facilitate important investigations, it also provides a
mechanism by which authorities can browse comprehensive or near-comprehensive
financial information about individuals without their knowledge.  Authorities
with those capabilities, and the businesses positioned to aggregate and analyse
data collected for compliance purposes, may also be able to conduct statistical
evaluations of individuals based upon the information available to their
financial institutions.  Once aggregated and linked to unitary identities, the
transaction data collected by financial institutions offer a detailed look into
the habits, patterns, travels, associations, and financial health of
individuals.

The risks associated with such surveillance of electronic transactions were
recognised fifty years ago by Paul Armer of the RAND Corporation, who
identified the risk in a 1968 US Senate deposition~\cite{armer1968} and later
argued that ``if you wanted to build an unobtrusive system for surveillance,
you couldn't do much better than an [electronic funds transfer
system]''~\cite{armer1975}.  Indeed, payment networks routinely share
information about financial transactions with credit bureaus such as
Experian~\cite{experian2018}, who are in the business of judging individuals by
their behaviours and whose judgments form the basis of decisions made by
lenders, insurers, and other clients of analytics companies~\cite{christl2017}.
Additionally, documents released by Edward Snowden have revealed that the US
National Security Agency has a division called ``Follow the Money'' (FTM) that
systematically collects and analyses data from payment
networks~\cite{spiegel2013}.

\subsection{Cryptocurrencies}

Cryptocurrencies have enjoyed popularity in recent years, and people have
flocked to cryptocurrencies for a variety of reasons.  The idea of accountless
digital cash is hardly new, dating at least as far back as the 1982 paper by
David Chaum on blind signatures~\cite{chaum1982}, the technology that he later
used to start DigiCash Inc, which folded in 1999~\cite{pitta1999}.  Other
attempts to develop accountless electronic payment systems such as
E-Gold~\cite{meek2007} and Liberty Reserve~\cite{bbc2013} were designed with
privacy in mind, and ultimately ran afoul of authorities when criminals used
those systems for nefarious purposes.

By the time Bitcoin emerged in early 2009~\cite{bitcoin}, the financial crisis
had prompted aggressive responses from central banks around the world, and
surely it was no coincidence that the message of circumventing inflationary
monetary policy enjoyed appeal among would-be hoarders.  However, given the
history of privacy as a primary motivation for the adoption of digital cash, we
surmise that many of the cryptocurrency adopters (other than speculators) are
primarily seeking privacy, whether to circumvent capital controls or just to
avoid the ``pastoral gaze'' of state or corporate
surveillance~\cite{sotirakopoulos2017}.  Some important developments in recent
years corroborate this view, most notably the attempts to develop a
``stablecoin'': a cryptocurrency that avoids the volatility of cryptocurrency
prices by establishing a market peg, for example to a fiat
currency~\cite{buterin2014}.  The most notorious example of a stablecoin is
Tether, a cryptocurrency that was established for the purpose of maintaining a
one-to-one peg with the US Dollar~\cite{popper2017,williams2018}.  For this
reason, stablecoins can be denominated in units of fiat currency.  However,
stablecoins have important limitations, including well-justified concerns about
unilateral exchange rate pegs in general~\cite{rogoff1998}.

As a replacement for the ``legitimate'' currencies underwritten by the full
faith and credit of sovereign governments, cryptocurrencies are far from
perfect.  There are structural reasons for this, including:

\begin{enumerate}

\item \textit{Absence of financial services.} There is a notable absence of
reliable organisations that offer routine financial services such as lending,
and more importantly, there is a lack of regulatory support for
crytpocurrencies.  Further, in contrast to transactions conducted via global
messaging systems such as SWIFT~\cite{swift}, there is generally no way to
correct or unwind erroneous transactions performed with permissionless
cryptocurrencies, a critical operational limitation.  For cryptocurrencies to
be a true substitute for government-issued currencies, they must support a
range of marketplaces and financial products.

\item \textit{Absence of regulated marketplaces.} History tells us that
unregulated marketplaces for financial products can be harmful to ordinary
citizens and businesses alike; consider for example the misbehaviour of brokers
and market participants that led to the creation of the US Securities and
Exchange Commission~\cite{sec-history}.  Cryptocurrency markets lack such
controls and mechanisms to ensure accountability, and unchecked market
manipulation is commonplace~\cite{tam2017,williams2017}.

\item \textit{Absence of legal context.} There is no generally applicable
mechanism for adjudicating disputes arising from transactions that are executed
in cryptocurrency.  When automatically executable contracts such as those that
underpinned the ``Decentralised Autonomous Organisation'' that roiled the
Ethereum community in 2016~\cite{falkon2017} are exploited, there is little
legal recourse for hapless victims.  Although ``certain operational clauses in
legal contracts'' may be automated to beneficial effect~\cite{isda2017}, it
would seem that a maximalist conception of the principle of ``code is law'' may
not be workable without a suitable legal framework.

\end{enumerate}

Furthermore, cryptocurrencies often fail to deliver on their key promises.  For
example, they are often not as private as is commonly believed.  Analysis of
Bitcoin transactions can deanonymise them, and researchers have shown that it
is eminently possible to identify meaningful patterns among the
transactions~\cite{meiklejohn2013,tasca2016}.  The problem persists not only as
a result of prevalent web trackers and the reuse of pseudonyms linked to
Bitcoin wallets~\cite{goldfeder2017} but also because inbound transactions to a
Bitcoin address can fundamentally be linked to outbound transactions from that
address~\cite{aljawaheri2018}.  Indeed, it has even been argued that the
explicit traceability of transactions on the Bitcoin ledger, combined with a
straightforward approach to tagging suspect transactions~\cite{anderson2018},
make it even less private than traditional mechanisms of payment.  Even
cryptocurrencies such as Monero, which are designed for privacy, have been
shown to have important weaknesses~\cite{moser2018,kappos2018}.  Another,
perhaps equally important deficiency of cryptocurrencies is that they are not
as decentralised as is commonly believed.  Although decentralisation is often
touted as the raison d'\^etre of cryptocurrencies~\cite{buterin2017}, in
practice the governance, ``mining,'' and infrastructure services associated
with cryptocurrencies have remained stubbornly centralised for a variety of
reasons~\cite{chepurnoy2017}.  The problem of decentralisation is intimately
related to the more elemental governance problem how to ensure that the system
serves the interest of its users.  Without institutional support, there is
little to ensure that this remains the case.

The governance problem is of particular significance to stablecoins.
Importantly, if a stablecoin is not maintained and controlled by a central
bank, then its users would need to be concerned about who is ultimately
providing assurance that it will retain its value.

\section{Electronic Payments Today}

As electronic funds transfer systems have proliferated in recent decades, so
has the expectation that people will make use of those systems.  If this trend
were to continue, we would anticipate that the fixed costs associated with
infrastructure to support cash transactions would become harder to justify, and
variable costs such as widespread deployment of ATM machines would be reduced.
Individuals and small businesses have various options to conduct transactions
electronically.  ``Electronic'' financial transactions include nearly all
economic transactions that are not conducted using cash, notwithstanding the
use of precious metals, money orders, and barter.  For our purposes, all
payments involving institutional accounts, including card payments (via payment
networks), wire transfers, ACH, and even physical cheques, are conducted
electronically, as are payments conducted using cryptocurrency.  Next, we shall
consider the characteristics of transactions in three examples of electronic
payments: those involving institutional accounts, those involving ``basic''
cryptocurrencies, and those involving ``privacy-enabling'' cryptocurrencies.

\subsection{Modern Retail Banking}

\begin{figure}
\begin{center}
\begin{tikzpicture}[>=latex, node distance=3cm, font={\sf \small}, auto]\ts
\node (r1) at (0,0) [noshape, text width=4em] {
    \scalebox{0.8}{\includegraphics{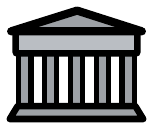}}
};
\node (r2) at (4,0) [noshape, text width=4em] {
    \scalebox{0.8}{\includegraphics{images/primary_bank.pdf}}
};
\node (r3) at (8,0) [noshape, text width=4em] {
    \scalebox{0.8}{\includegraphics{images/primary_bank.pdf}}
};
\node (m1) at (0,1) [noshape] {
    \scalebox{0.06}{\includegraphics{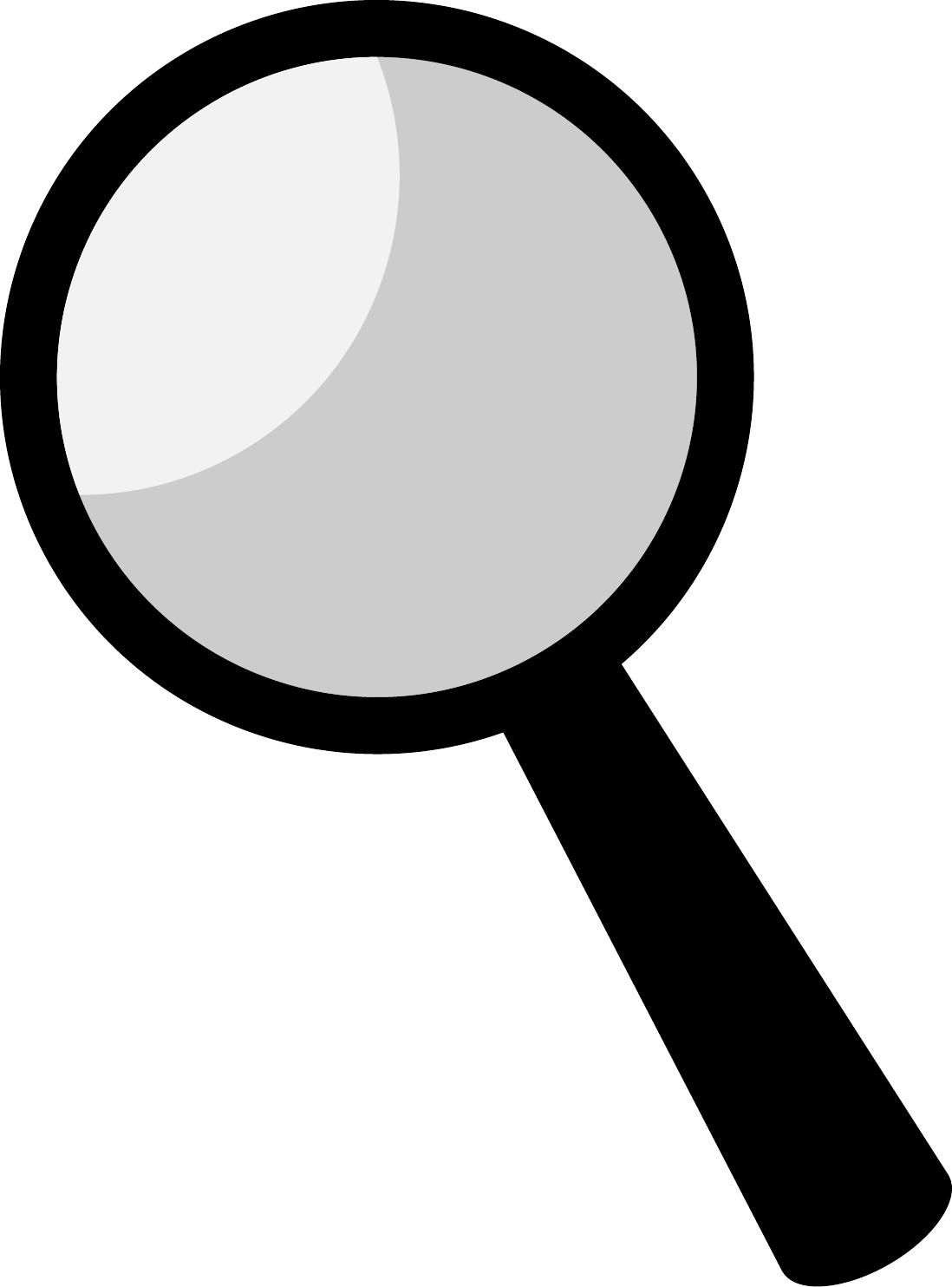}}
};
\node (m2) at (4,1) [noshape] {
    \scalebox{0.06}{\includegraphics{images/office-glass-magnify.pdf}}
};
\node (m3) at (8,1) [noshape] {
    \scalebox{0.06}{\includegraphics{images/office-glass-magnify.pdf}}
};

\draw[->, line width=0.5mm] (r1) -- node[below] {\textbf{\textit{\Huge \pounds}}} node[above] {
    \scalebox{0.06}{\includegraphics{images/office-glass-magnify.pdf}}
} (r2);
\draw[->, line width=0.5mm] (r2) -- node[below] {\textbf{\textit{\Huge \pounds}}} node[above] {
    \scalebox{0.06}{\includegraphics{images/office-glass-magnify.pdf}}
} (r3);

\end{tikzpicture}

\caption{\textit{Schematic Representation of Modern Retail Banking Transaction
Flows.}  The buildings with columns represent financial institutions with which
the transacting parties hold accounts.  Money is exchanged in state-issued
currency, as represented by the Pound Sterling symbols.  Authorities and other
powerful actors can monitor both the institutions and the flows, as represented
by the magnifying glasses.}

\label{f:mrb}
\end{center}
\end{figure}

Modern retail banking involves electronic transactions between
\textit{accounts}, each of which represents a bilateral relationship between a
financial \textit{institution} (e.g., a bank) and another entity, perhaps an
individual.  Institutions are generally regulated by governments.  Individuals
and businesses may agree to exchange value (for example, in return for goods
and services), but in reality the transaction takes place between institutions,
which mutually agree to modify the state of the accounts such that the account
of the ``receiver'' is incremented and the account of the ``sender'' is
decremented correspondingly.  Record of the transactions and their results are
generally visible to the institutions, accountholders, authorities, and
auditors.  Figure~\ref{f:mrb} offers an illustration of the data flows
corresponding to two transactions.  Institutions are direct participants in the
transactions.  Both the accounts and the transactions may be monitored, i.e.
``external'' \textit{observers} such as authorities (and in some cases others,
such as unprivileged employees of the institutions and hackers) are able to
examine the records of the transactions, their results, and the transactions
themselves.  Since the set of regulated institutions is small, it is efficient
for an observer to collect, aggregate, and analyse the data associated with
substantially all of the transactions that take place within the system.

By contrast, data on transactions involving cash are relatively difficult to
observe in this fashion, and are therefore more private.  However, although
cash remains a popular instrument for retail transactions, its use is
decreasing as consumers become more comfortable with electronic means of
payment~\cite{matheny2016}.  Some economists such as Kenneth Rogoff hail this
transformation as a welcome development, citing reductions in tax evasion and
crime as primary benefits as anonymous payments are
curtailed~\cite{rogoff2017}.  Others are more circumspect.  Citing Sweden's
drive to become cashless, Jonas Hedman recognised the loss of privacy as the
primary disadvantage of a cashless society, although he also acknowledged that
the transition to cashlessness is inevitable~\cite{wharton2018}.  Assuming that
the insistence on unitary identifiers for all electronic financial transactions
as proposed by regulations such as 5AMLD~\cite{eu5amld} is satisfied, and
combined with large-scale aggregation and analysis of the sort already in
practice~\cite{spiegel2013}, cashlessness means the creation of a browseable
``permanent record'' for every individual containing his or her entire
transaction history.

\subsection{``Basic'' Cryptocurrency, e.g. Bitcoin}

\begin{figure}
\begin{center}
\begin{tikzpicture}[>=latex, node distance=3cm, font={\sf \small}, auto]\ts
\node (r1) at (0,0) [noshape, text width=4em] {
    \scalebox{1.2}{\includegraphics{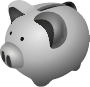}}
};
\node (r2) at (4,0) [noshape, text width=4em] {
    \scalebox{1.2}{\includegraphics{images/gray_piggybank.pdf}}
};
\node (r3) at (8,0) [noshape, text width=4em] {
    \scalebox{1.2}{\includegraphics{images/gray_piggybank.pdf}}
};
\node (m1) at (0,1.1) [noshape] {
    \scalebox{0.06}{\includegraphics{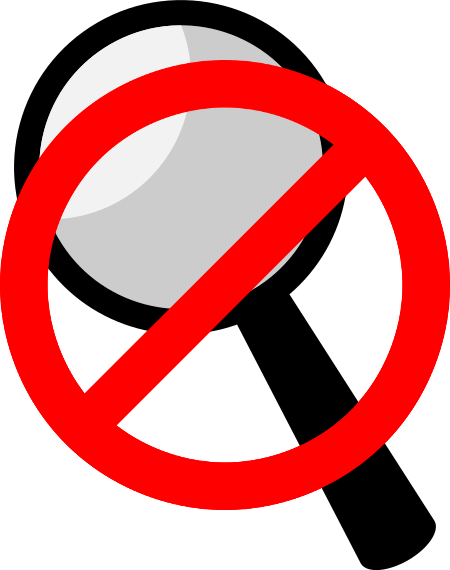}}
};
\node (m2) at (4,1.1) [noshape] {
    \scalebox{0.06}{\includegraphics{images/office-glass-magnify-no.png}}
};
\node (m3) at (8,1.1) [noshape] {
    \scalebox{0.06}{\includegraphics{images/office-glass-magnify-no.png}}
};

\draw[->, line width=0.5mm] (r1) -- node[below] {
    \scalebox{0.6}{\includegraphics{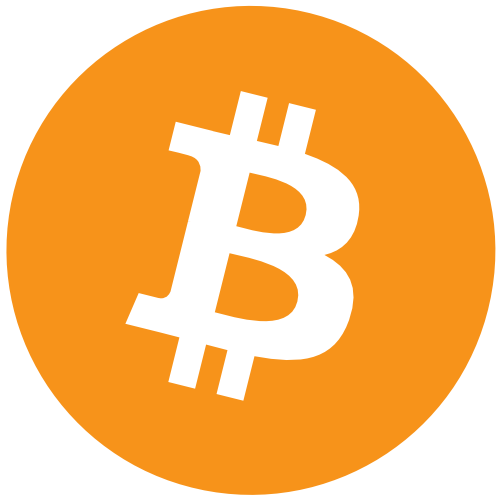}}
} node[above] {
    \scalebox{0.06}{\includegraphics{images/office-glass-magnify.pdf}}
} (r2);
\draw[->, line width=0.5mm] (r2) -- node[below] {
    \scalebox{0.6}{\includegraphics{images/bitcoin.png}}
} node[above] {
    \scalebox{0.06}{\includegraphics{images/office-glass-magnify.pdf}}
} (r3);

\end{tikzpicture}

\caption{\textit{Schematic Representation of Bitcoin Transaction Flows.}
Transacting parties can store value on their own devices, represented as piggy
banks.  The flows can be monitored by anyone.}

\label{f:bitcoin}
\end{center}
\end{figure}

Cryptocurrencies offer an alternative payment mechanism that avoids some
aspects of the surveillance infrastructure that characterises
institutionally-mediated retail bank transactions.  Modern cryptocurrencies
generally take the form of bearer instruments, in the sense that their units
are each represented by a public key on a public ledger and controlled by the
knowledge of the matching private key.  Users are not required to establish
accounts or furnish identification information of any sort to receive, possess,
or spend cryptocurrency.  This is not to say that accounts do not exist; most
users of popular cryptocurrencies such as Bitcoin and Ethereum establish
accounts with centralised wallet providers such as \textit{blockchain.info} or
\textit{myetherwallet}~\cite{chepurnoy2017}.  Providers of accounts could be
compromised or subverted by state actors or other powerful groups with an
interest in surveillance.  Some account platforms cooperate with national
regulators~\cite{sec2018}, and some national regulators have declared that they
will limit the scope of the rules that would apply to such
platforms~\cite{higgins2016}.  Many if not most cryptocurrency transactions are
done by speculators, not those who intend to use cryptocurrency for its
fundamental properties~\cite{russo2018}, so even if most traders in practice
might be indifferent to strong identity requirements crafted by regulators to
satisfy AML goals, such rules undermine a key design objective of
cryptocurrencies themselves.

In principle, however, users of cryptocurrencies are not required to register
with platforms, and they may possess cryptocurrency tokens on their own
devices.  Figure~\ref{f:bitcoin} shows how this works in practice.  Assuming
that cryptocurrency users take precautions not to reveal their identities
whilst transacting, for example by using anonymity systems such as
Tor~\cite{tor}, they might expect to avoid identity-based blacklisting when
they receive tokens.  However, depending upon the system design, adversaries
may still be able to monitor the flows.  Because successive Bitcoin
transactions are linkable to each other, those able to monitor the network can
determine successive transactions associated with specific tokens and
ultimately deanonymise Bitcoin
users~\cite{meiklejohn2013,tasca2016,aljawaheri2018}.

The fact that individual tokens can be traced means that cryptocurrencies such
as Bitcoin may not be entirely fungible, in the sense of being ``easy to
exchange or trade for something else of the same type and
value''~\cite{cam-fungible}, as an individual might be less willing to accept
certain specific cryptocurrency tokens because doing so might implicitly link
that individual to previous owners of the tokens.  Traceability has created
demand for newly-minted or ``clean'' tokens that are harder to link to the
previous owners or (ultimately) the previous transactions of the current
owner~\cite{osborne2016}, and the proposed blacklists of cryptocurrency
addresses associated with suspicious operators could further exacerbate this
distinction~\cite{hinkes2018}.  To avoid this problem, a cryptocurrency
implementation would need to offer assurance that a transaction by an asset
holder would generally not, directly or indirectly, result in that asset holder
being linked to other transactions that had taken place previously.
Additionally, cryptocurrencies that make use of immutable ledgers and do not
protect against traceability may for that reason be non-compliant with data
protection regulations such as GDPR that specify a ``right to be
forgotten''~\cite{maxwell2017}.

\subsection{``Privacy-Enabling'' Cryptocurrency, e.g. Monero}
\label{ss:pec}

\begin{figure}
\begin{center}
\begin{tikzpicture}[>=latex, node distance=3cm, font={\sf \small}, auto]\ts
\node (r1) at (0,0) [noshape, text width=4em] {
    \scalebox{1.2}{\includegraphics{images/gray_piggybank.pdf}}
};
\node (r2) at (4,0) [noshape, text width=4em] {
    \scalebox{1.2}{\includegraphics{images/gray_piggybank.pdf}}
};
\node (r3) at (8,0) [noshape, text width=4em] {
    \scalebox{1.2}{\includegraphics{images/gray_piggybank.pdf}}
};
\node (m1) at (0,1.1) [noshape] {
    \scalebox{0.06}{\includegraphics{images/office-glass-magnify-no.png}}
};
\node (m2) at (4,1.1) [noshape] {
    \scalebox{0.06}{\includegraphics{images/office-glass-magnify-no.png}}
};
\node (m3) at (8,1.1) [noshape] {
    \scalebox{0.06}{\includegraphics{images/office-glass-magnify-no.png}}
};

\draw[->, line width=0.5mm] (r1) -- node[below] {
    \scalebox{0.03}{\includegraphics{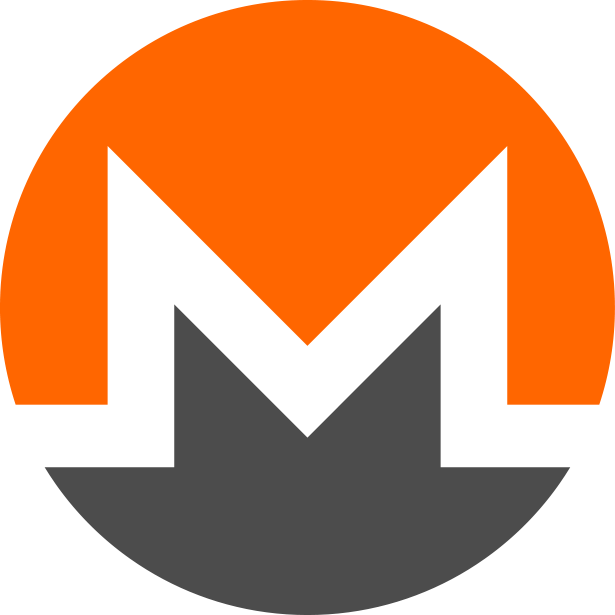}}
} node[above] {
    \scalebox{0.06}{\includegraphics{images/office-glass-magnify-no.png}}
} (r2);
\draw[->, line width=0.5mm] (r2) -- node[below] {
    \scalebox{0.03}{\includegraphics{images/monero.png}}
} node[above] {
    \scalebox{0.06}{\includegraphics{images/office-glass-magnify-no.png}}
} (r3);

\end{tikzpicture}

\caption{\textit{Schematic Representation of Monero Transaction Flows.}  In an
``idealised'' version of Monero or other privacy-enabling cryptocurrency,
observers would not be able to infer information about transacting parties or
the relationships between transactions by monitoring the ledger or the
transactions themselves, as indicated by the magnifying glasses with the
negation symbol.  The piggy banks indicate that users are storing the tokens
privately rather than relying upon accounts.}

\label{f:monero}
\end{center}
\end{figure}

Some cryptocurrencies, most notably Zcash and Monero, are explicitly designed
to address traceability concerns~\cite{sedgwick2018}.  Monero in particular
takes an approach that incorporates several security mechanisms, including:

\begin{enumerate}

\item \textit{Ring Signatures,} which allow signed messages to be attributable
to ``a set of possible signers without revealing which member actually produced
the signature''~\cite{rivest2001}.

\item \textit{Stealth Addresses,} which refer to methods for key management in
which public keys are derived separately from private keys for the purpose of
obscuring the public keys~\cite{courtois2017}, and

\item \textit{Confidential Transactions,} which use Pedersen commitment
schemes~\cite{pedersen1991} to restrict disclosing the amounts transacted to
anyone other than the transacting parties~\cite{vanwirdum2016}.

\end{enumerate}

Figure~\ref{f:monero} illustrates how, in a successfully implemented
privacy-enabling cryptocurrency, metadata associated with transactions would be
hidden such that the data flows or the ledger would not reveal relationships
among transactions or any information about the transacting parties.  That
said, the Monero design and implementation still do not completely realise this
goal; its process for mixing transactions suffers from inconsistent selection
probability among all elements of the anonymity set~\cite{moser2018}.  Monero
spokesperson Riccardo Spagni countered that ``privacy isn’t a thing you
achieve, it’s a constant cat-and-mouse battle''~\cite{greenberg2018}, echoing
longstanding arguments by others that privacy is inevitably an endeavour of
vigilance and responsiveness~\cite{zimmermann1991}.

Some authorities such as the Japanese Financial Security Agency
(FSA)~\cite{wilmoth2018,viglione2018} and the United States Secret
Service~\cite{novy2018} have responded to so-called ``privacy coins'' by
banning the use of privacy-enhancing cryptocurrencies whilst accepting other
cryptocurrencies as legitimate by comparison.  For a cryptocurrency exchange or
other provider of cryptocurrency-based financial services to be compliant under
such rules, it would need to restrict its activities to cryptocurrencies such
as Bitcoin and Ethereum which do not have the privacy characteristics that have
been sought by cryptocurrency advocates for decades.

There have also been some attempts, notably Mimblewimble~\cite{jedusor2016}, to
retrofit basic cryptocurrencies with some of the characteristics of
privacy-enabling cryptocurrencies, although it remains to be seen whether such
approaches will turn out to be more effective than cryptocurrencies designed
with better intrinsic privacy features in the first instance.






\section{Proposed Hybrid Approaches}

\begin{table}
\begin{center}
\sf\begin{tabular}{|L{4cm}|p{1.2cm}p{1.2cm}p{1.2cm}p{1.2cm}p{1.2cm}|}\hline
& \rotatebox{90}{cash}
& \rotatebox{90}{modern} \rotatebox{90}{retail banking}
& \rotatebox{90}{``traditional''} \rotatebox{90}{cryptocurrency} \rotatebox{90}{(e.g. Bitcoin)}
& \rotatebox{90}{``traditional''} \rotatebox{90}{stablecoins} \rotatebox{90}{(e.g. Tether)}
& \rotatebox{90}{privacy-enabling\hspace{8pt}} \rotatebox{90}{cryptocurrency} \rotatebox{90}{(e.g. Monero)} \\
Robust to cyberattacks      & \CIRCLE & \Circle & \Circle & \Circle & \Circle \\
Usable without registration & \CIRCLE & \Circle & \CIRCLE & \CIRCLE & \CIRCLE \\
Unlinkable* transactions    & \RIGHTcircle & \Circle & \Circle & \Circle & \CIRCLE \\
Electronic transactions     & \Circle & \CIRCLE & \CIRCLE & \CIRCLE & \CIRCLE \\
Suitable for taxation       & \RIGHTcircle & \CIRCLE & \Circle & \RIGHTcircle & \Circle \\
Can block some illicit uses & \Circle & \CIRCLE & \Circle & \Circle & \Circle \\
Can be denominated in       & \CIRCLE & \CIRCLE & \Circle & \CIRCLE & \Circle \\
\vspace{-4.5mm}\hspace{3mm} units of fiat currency & & & & & \\
\hline\end{tabular}\rm
\vspace{-6pt}
\end{center}

\sf\small\hspace{56pt}*Potentially\rm

\caption{\textit{Comparison of various existing electronic payment methods.} [*Potentially]}

\label{t:existing}
\end{table}

We consider the following challenge facing policymakers, regulators, and
technologists alike: \textit{how can we achieve realise the benefits of
government regulation without creating a central database that irreversibly
connects all persons with all of their transactions?}  There are two parts to
this question.  The first part is primarily about \textit{technology:} can we
build a system that securely processes financial transactions conducted
electronically without revealing data about the transacting parties?  The
answer is yes, as described in our discussion of privacy-enabling
cryptocurrency, with an important qualification that privacy is really an
iterative process that can only really be developed through active commitment
and ongoing vigilance.  The second part is primarily about \textit{government
policy:}

\begin{itemize}

\item What exactly are the key government objectives for regulating
transactions?

\item Which objectives are essential, and which can be deprioritised?

\item Do any of the objectives conflict with the human right to privacy?

\end{itemize}

Table~\ref{t:existing} shows how the existing payment methods achieve the
desiderata listed in Section~\ref{s:intro}.  (None of the popular
cryptocurrencies are known to offer totally unlinkable transactions, continual
improvements notwithstanding.)  Can we achieve a compromise that does better
than the prevailing methods for electronic payments?

In this section we introduce two approaches to frame the discussion of how to
resolve the tension.  The first approach, \textit{institutionally supported
privacy-enabling cryptocurrency}, provides regulated institutions with tools
and procedures for interacting with privacy-enabling cryptocurrencies, creating
a structure for legal interpretations of their use.  We assume that the
distributed ledgers underlying such cryptocurrencies \textit{are not}
controlled by regulated financial institutions.  The second approach,
\textit{institutionally mediated private value exchange}, establishes a method
by which regulated institutions can conduct financial transactions on a
distributed ledger that shares essential characteristics with privacy-enabling
cryptocurrencies.  In this case, we assume that the distributed ledgers used
for this purpose \textit{are} controlled by regulated financial institutions.
The main difference between the two approaches is that the first approach
allows businesses to transact with cryptocurrencies that are managed and
governed outside the mainstream financial system, and the second approach
provides a way for regulated financial institutions to offer a mechanism for
their clients to exchange money that resembles cryptocurrency in that clients
can withdraw money electronically and subsequently use it without reference to
an account, as they would with cash.

There is a third possibility, which we might describe as ``institutionally
supported privacy-enabling stablecoins,'' in which the privacy-enabling
cryptocurrencies in question are actually stablecoins.  This possibility is
theoretically worth pursuing if stablecoins achieve popularity commensurate
with cryptocurrencies, although the experience of Tether suggests it might not
be easy.  It is worth considering that the proposal for institutionally
mediated private value exchange is similar to a stablecoin in that the tokens
represent units of fiat currency.  However, because the regulated financial
institutions are assumed to be part of the banking system they would not need
to bear the risk associated with maintaining a market peg.

\subsection{Institutionally Supported Privacy-Enabling Cryptocurrency}
\label{ss:isc}

\begin{figure}
\begin{center}
\begin{tikzpicture}[>=latex, node distance=3cm, font={\sf \small}, auto]\ts
\node (r1) at (0.0,0) [noshape, text width=4em] {
    \scalebox{0.8}{\includegraphics{images/primary_bank.pdf}}
};
\node (m2) at (-1.0,0) [noshape] {
    \scalebox{0.06}{\includegraphics{images/office-glass-magnify.pdf}}
};
\node (r2) at (1.6,2.5) [noshape, text width=4em] {
    \scalebox{0.14}{\includegraphics{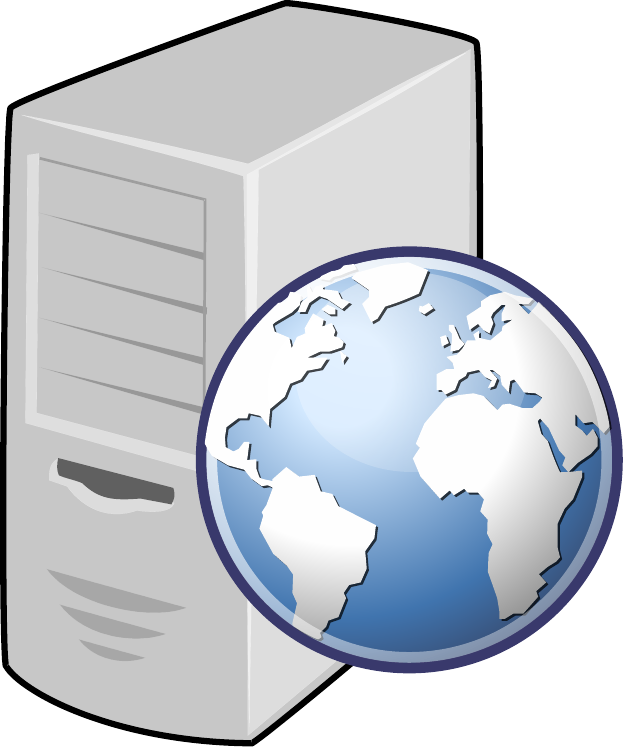}}
};
\node (m2) at (0.6,2.5) [noshape] {
    \scalebox{0.06}{\includegraphics{images/office-glass-magnify-no.png}}
};
\node (r3) at (1.6,-2.5) [noshape, text width=4em] {
    \scalebox{0.14}{\includegraphics{images/web_server.pdf}}
};
\node (m2) at (0.6,-2.5) [noshape] {
    \scalebox{0.06}{\includegraphics{images/office-glass-magnify-no.png}}
};
\node (r4) at (4.4,2.5) [noshape, text width=4em] {
    \scalebox{0.8}{\includegraphics{images/primary_bank.pdf}}
};
\node (m2) at (5.4,2.5) [noshape] {
    \scalebox{0.06}{\includegraphics{images/office-glass-magnify.pdf}}
};
\node (r5) at (4.4,-2.5) [noshape, text width=4em] {
    \scalebox{0.8}{\includegraphics{images/primary_bank.pdf}}
};
\node (m2) at (5.4,-2.5) [noshape] {
    \scalebox{0.06}{\includegraphics{images/office-glass-magnify.pdf}}
};
\node (r6) at (6.0,0) [noshape, text width=4em] {
    \scalebox{0.14}{\includegraphics{images/web_server.pdf}}
};
\node (m2) at (7.0,0) [noshape] {
    \scalebox{0.06}{\includegraphics{images/office-glass-magnify-no.png}}
};

\draw[-, line width=0.5mm] (r1) edge[above,bend left=15] node[above,xshift=-15] {
} (r2);
\draw[-, line width=0.5mm] (r2) edge[above,bend left=15] node[above] {
} (r4);
\draw[-, line width=0.5mm] (r4) edge[above,bend left=15] node[above,xshift=15] {
} (r6);
\draw[-, line width=0.5mm] (r6) edge[above,bend left=15] node[above,xshift=15] {
} (r5);
\draw[-, line width=0.5mm] (r5) edge[above,bend left=15] node[above,xshift=15] {
} (r3);
\draw[-, line width=0.5mm] (r3) edge[above,bend left=15] node[above,xshift=15] {
} (r1);

\end{tikzpicture}

\caption{\textit{Schematic Representation of Institutionally Supported
Privacy-Enabling Cryptocurrency: Nodes.} Institutions would join global
networks of servers operating as nodes in existing cryptocurrency networks; not
all participants in these networks are regulated institutions.}

\label{f:an}
\end{center}
\end{figure}

\begin{figure}
\begin{center}
\begin{tikzpicture}[>=latex, node distance=3cm, font={\sf \small}, auto]\ts
\node (r1) at (0,0) [noshape, text width=4em] {
    \scalebox{0.8}{\includegraphics{images/primary_bank.pdf}}
};
\node (o1) at (0.4,-0.3) [noshape, text width=4em] {
    \scalebox{0.9}{\includegraphics{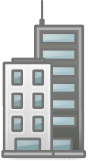}}
};
\node (r2) at (4,0) [noshape, text width=4em] {
    \scalebox{0.8}{\includegraphics{images/primary_bank.pdf}}
};
\node (o2) at (4.4,-0.4) [noshape, text width=4em] {
    \scalebox{0.4}{\includegraphics{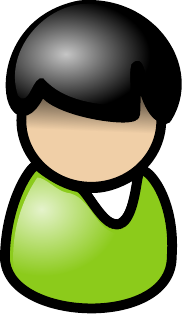}}
};
\node (r3) at (8,0) [noshape, text width=4em] {
    \scalebox{1.2}{\includegraphics{images/gray_piggybank.pdf}}
};
\node (o3) at (8.4,-0.4) [noshape, text width=4em] {
    \scalebox{0.4}{\includegraphics{images/people-juliane-krug-08a.pdf}}
};
\node (m1) at (0,1) [noshape] {
    \scalebox{0.06}{\includegraphics{images/office-glass-magnify.pdf}}
};
\node (m2) at (4,1) [noshape] {
    \scalebox{0.06}{\includegraphics{images/office-glass-magnify.pdf}}
};
\node (m3) at (8,1.1) [noshape] {
    \scalebox{0.06}{\includegraphics{images/office-glass-magnify-no.png}}
};

\draw[->, line width=0.5mm] (r1) -- node[below] {
    \scalebox{0.03}{\includegraphics{images/monero.png}}
} node[above] {
} (r2);
\draw[->, line width=0.5mm] (r2) -- node[below] {
    \scalebox{0.03}{\includegraphics{images/monero.png}}
} node[above] {
    \scalebox{0.06}{\includegraphics{images/office-glass-magnify-no.png}}
} (r3);

\end{tikzpicture}

\caption{\textit{Schematic Representation of Institutionally Supported
Privacy-Enabling Cryptocurrency: Transaction Flows (1).}  (We use the Monero
symbol to represent any privacy-enabling cryptocurrency without loss of
generality.) Corporations and registered businesses with accounts held by
regulated financial institutions (leftmost icon) that would be subject to
monitoring and may only remit cryptocurrency payments to other accounts held by
regulated financial institutions.  Individuals and non-business partnerships
(centre icon) may transfer cryptocurrency from accounts to unmonitored, private
storage (rightmost icon).}

\label{f:atf1}
\end{center}
\end{figure}

Our first approach starts with existing, privacy-enabling cryptocurrencies such
as Zcash or Monero and assumes that regulators have chosen to embrace the new
methods for exchanging value and accept, if not support outright, at least some
of the various communities that have formed around particular cryptocurrencies
to provide governance and software development.  Acceptance of cryptocurrencies
by governments and other institutions is certainly plausible; for instance, the
Bank of England concluded that cryptocurrencies ``currently do not pose a
material risk to UK financial stability''~\cite{boe2018}.  It assumes that
government priorities include collecting taxes and monitoring transactions
undertaken by businesses and regulated institutions.

Figure~\ref{f:an} illustrates how institutions would join existing
cryptocurrency systems as full participants.  The motivation for broker-dealers
and other institutions to participate is well-established; financial services
related to cryptocurrencies are in demand by hedge funds and other
clients~\cite{verhage2018,hankin2018}.  Of course, this implies that
broker-dealers would likely undertake activities related to unregulated markets
and marketplaces (i.e., the cryptocurrencies themselves), and presumably the
governance of the cryptocurrencies would not be under institutional control.
That said, the distributed ledger underlying the cryptocurrencies would ensure
that there would be an audit trail of all transactions, even if the details of
those transactions might be inscrutable to authorities, auditors, or others
without the active participation of the transacting parties.

\begin{figure}
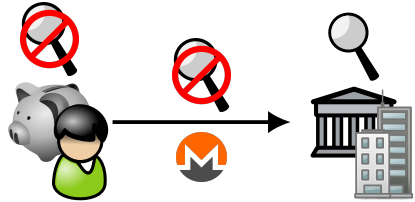

\begin{center}
\begin{tikzpicture}[>=latex, node distance=3cm, font={\sf \small}, auto]\ts
\node (r1) at (0,0) [noshape, text width=4em] {
    \scalebox{1.2}{\includegraphics{images/gray_piggybank.pdf}}
};
\node (o2) at (0.4,-0.4) [noshape, text width=4em] {
    \scalebox{0.4}{\includegraphics{images/people-juliane-krug-08a.pdf}}
};
\node (r2) at (4,0) [noshape, text width=4em] {
    \scalebox{0.8}{\includegraphics{images/primary_bank.pdf}}
};
\node (o1) at (4.4,-0.3) [noshape, text width=4em] {
    \scalebox{0.9}{\includegraphics{images/office-towers.pdf}}
};
\node (m2) at (4,1) [noshape] {
    \scalebox{0.06}{\includegraphics{images/office-glass-magnify.pdf}}
};
\node (m1) at (0,1.1) [noshape] {
    \scalebox{0.06}{\includegraphics{images/office-glass-magnify-no.png}}
};

\draw[->, line width=0.5mm] (r1) -- node[below] {
    \scalebox{0.03}{\includegraphics{images/monero.png}}
} node[above] {
    \scalebox{0.06}{\includegraphics{images/office-glass-magnify-no.png}}
} (r2);

\end{tikzpicture}

\caption{\textit{Schematic Representation of Institutionally Supported
Privacy-Enabling Cryptocurrency: Transaction Flows (2).}  An individual (shown
at left) with a private store of cryptocurrency could remit payments without
revealing her identity to a business with accounts held by a regulated
institution (shown at right).}

\label{f:atf2}
\end{center}
\end{figure}

To facilitate monitoring, auditing, and taxation, we assume that regulators
would stipulate that all cryptocurrency transactions undertaken by certain
legal entities other than individual persons, such as registered corporations,
licensed businesses, charities, trusts, and some partnerships, must take place
via regulated institutional intermediaries such as banks, custodians, or
broker-dealers.  In general, such legal entities are already subject to various
forms of government oversight, for example tax reporting requirements, so to
introduce additional requirements and enforceability for cryptocurrency
transactions is not unfathomable.  The institutions would carry out AML/KYC
compliance procedures as they currently do, and regulators would require that
all cryptocurrency disbursements from such registered corporations or
organisations, including dividends, interest, proceeds from disposal of
cryptocurrency-denominated assets, and payments, including without limitation
payments to suppliers, service providers, employees, and contractors, would
take the form of remittances to other institutional accounts that hold
cryptocurrency.

Individuals and non-business partnerships would not be subject to the same
requirements and would be permitted to transact and hold cryptocurrency
privately, as they do in many countries today.  Figure~\ref{f:atf1} shows how
this would work in practice.  Businesses would maintain accounts with
institutions and could direct the institutions to remit payments to other
institutionally held accounts, including those whose beneficial owners are
individuals, and individuals could in turn direct their institutional accounts
to remit payments to their private cryptocurrency storage, which might or might
not be hosted by a wallet provider.  Individuals could then remit payments from
their own private storage to regulated businesses, such as merchants, private
organisations, or service providers, without necessarily revealing their
identities or a link to previous transactions such as those from which they
received the cryptocurrency in the first place; Figure~\ref{f:atf2} offers an
illustration.  Given that the legal entities covered in the last paragraph are
typically subject to financial reporting requirements, for example to quantify
reimbursements or to reconcile changes in assets with income, we assert that it
would be no easier for a business to deputise an individual to conduct
cryptocurrency transactions on its behalf than it would for a business to
deputise an individual to conduct any other financially meaningful aspect of
its business.

Dividing the different ways of holding the cryptocurrency into two categories
based upon whether or not it is held via accounts associated with regulated
institutions may be considered analogous to dividing Zcash into ``T''
(Transparent) and ``Z'' (Shielded) addresses~\cite{peterson2016}.

Because all cryptocurrency accounts held by corporations and registered
businesses would be subject to monitoring by regulated institutions, the
infrastructure would ensure that the taxable income of such corporations and
businesses would be known.  Because all payments from corporations and
registered businesses must be remitted to other institutional accounts, the
infrastructure would ensure that the income of their shareholders, suppliers,
service providers, and employees would be known and attributable to the correct
legal entities.  Authorities would realise other benefits as well.  The
distributed ledger maintained by the cryptocurrency node operators would be
observable by regulators and other authorities and cross-referenced against any
cash flow statements of businesses engaged in cryptocurrency transactions.
Private transactions suspected of criminal activity could be verified by
investigators with the cooperation of one of the counterparties, even if the
investigation might not necessarily reveal identifying details of the other
counterparty.

\begin{figure}
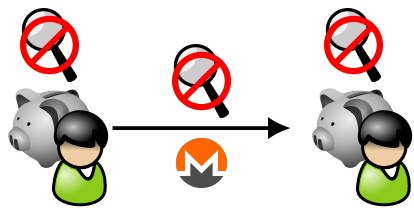

\begin{center}
\begin{tikzpicture}[>=latex, node distance=3cm, font={\sf \small}, auto]\ts
\node (r1) at (0,0) [noshape, text width=4em] {
    \scalebox{1.2}{\includegraphics{images/gray_piggybank.pdf}}
};
\node (o2) at (0.4,-0.4) [noshape, text width=4em] {
    \scalebox{0.4}{\includegraphics{images/people-juliane-krug-08a.pdf}}
};
\node (r2) at (4,0) [noshape, text width=4em] {
    \scalebox{1.2}{\includegraphics{images/gray_piggybank.pdf}}
};
\node (o2) at (4.4,-0.4) [noshape, text width=4em] {
    \scalebox{0.4}{\includegraphics{images/people-juliane-krug-08a.pdf}}
};
\node (m1) at (0,1.1) [noshape] {
    \scalebox{0.06}{\includegraphics{images/office-glass-magnify-no.png}}
};
\node (m2) at (4,1.1) [noshape] {
    \scalebox{0.06}{\includegraphics{images/office-glass-magnify-no.png}}
};

\draw[->, line width=0.5mm] (r1) -- node[below] {
    \scalebox{0.03}{\includegraphics{images/monero.png}}
} node[above] {
    \scalebox{0.06}{\includegraphics{images/office-glass-magnify-no.png}}
} (r2);

\end{tikzpicture}

\caption{\textit{Schematic Representation of Institutionally Supported
Privacy-Enabling Cryptocurrency: Transaction Flows (3).}  Individuals with
private stores of privacy-enabling cryptocurrency may transact directly without
revealing their identities.}

\label{f:atf3}
\end{center}
\end{figure}

One type of transaction under this system that might be of particular concern
to authorities is illustrated by Figure~\ref{f:atf3}, in which an individual
with a private cryptocurrency store remits cryptocurrency to another individual
with a private cryptocurrency store, not involving a regulated institution.
The fact that such transactions could take place without the involvement of
institutions means that authorities would be unable to completely enforce
restrictions on who is able to transact, in accordance with the FATF
recommendations~\cite{fatf-recommendations}.  We could argue that value will
find its way to criminal organisations with or without the sanctions advised by
FATF~\cite{cei1999}, or that those willing to break the law have many options
to anonymously acquire ``legitimate'' accounts~\cite{hern2015}, or that
prospective money launderers with sufficient assets will find other ways to
transact outside the the system.  Whether or not such arguments are sound,
cryptocurrencies might become a dominant form of exchanging value precisely
because people value privacy, in which case regulators will need to support
cryptocurrency transactions simply because those are the transactions that are
taking place.  After all, people certainly exchanged value before central banks
started issuing currency.

Another, equally important, characteristic of this approach is that without
institutional mediation at their core, cryptocurrencies are subject to the
vicissitudes of mining pools, hackers, and powerful global-scale actors who
might compromise or hijack them, as well as speculators and market manipulators
who might simply deplete their value.  However, an alternative interpretation
of that property is that different cryptocurrencies would compete with each
other, not only on the basis of market penetration but also on the basis of
privacy.  It is difficult to imagine a currency in a monopoly position,
state-sponsored or otherwise, having this characteristic.

\subsection{Institutionally Mediated Private Value Exchange}

\begin{figure}
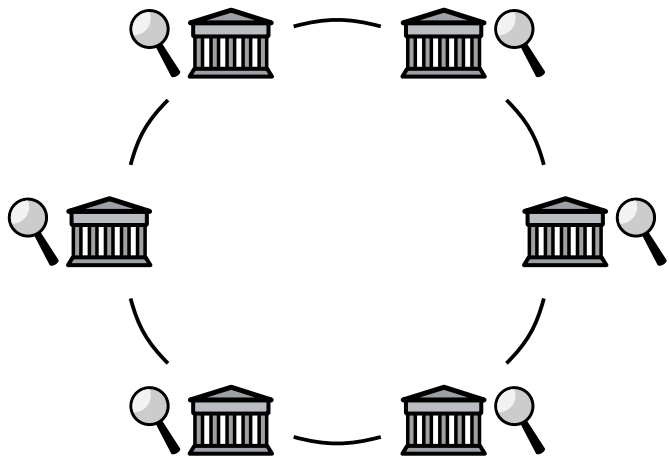

\begin{center}
\begin{tikzpicture}[>=latex, node distance=3cm, font={\sf \small}, auto]\ts
\node (r1) at (0.0,0) [noshape, text width=4em] {
    \scalebox{0.8}{\includegraphics{images/primary_bank.pdf}}
};
\node (m2) at (-1.0,0) [noshape] {
    \scalebox{0.06}{\includegraphics{images/office-glass-magnify.pdf}}
};
\node (r2) at (1.6,2.5) [noshape, text width=4em] {
    \scalebox{0.8}{\includegraphics{images/primary_bank.pdf}}
};
\node (m2) at (0.6,2.5) [noshape] {
    \scalebox{0.06}{\includegraphics{images/office-glass-magnify.pdf}}
};
\node (r3) at (1.6,-2.5) [noshape, text width=4em] {
    \scalebox{0.8}{\includegraphics{images/primary_bank.pdf}}
};
\node (m2) at (0.6,-2.5) [noshape] {
    \scalebox{0.06}{\includegraphics{images/office-glass-magnify.pdf}}
};
\node (r4) at (4.4,2.5) [noshape, text width=4em] {
    \scalebox{0.8}{\includegraphics{images/primary_bank.pdf}}
};
\node (m2) at (5.4,2.5) [noshape] {
    \scalebox{0.06}{\includegraphics{images/office-glass-magnify.pdf}}
};
\node (r5) at (4.4,-2.5) [noshape, text width=4em] {
    \scalebox{0.8}{\includegraphics{images/primary_bank.pdf}}
};
\node (m2) at (5.4,-2.5) [noshape] {
    \scalebox{0.06}{\includegraphics{images/office-glass-magnify.pdf}}
};
\node (r6) at (6.0,0) [noshape, text width=4em] {
    \scalebox{0.8}{\includegraphics{images/primary_bank.pdf}}
};
\node (m2) at (7.0,0) [noshape] {
    \scalebox{0.06}{\includegraphics{images/office-glass-magnify.pdf}}
};

\draw[-, line width=0.5mm] (r1) edge[above,bend left=15] node[above,xshift=-15] {
} (r2);
\draw[-, line width=0.5mm] (r2) edge[above,bend left=15] node[above] {
} (r4);
\draw[-, line width=0.5mm] (r4) edge[above,bend left=15] node[above,xshift=15] {
} (r6);
\draw[-, line width=0.5mm] (r6) edge[above,bend left=15] node[above,xshift=15] {
} (r5);
\draw[-, line width=0.5mm] (r5) edge[above,bend left=15] node[above,xshift=15] {
} (r3);
\draw[-, line width=0.5mm] (r3) edge[above,bend left=15] node[above,xshift=15] {
} (r1);

\end{tikzpicture}

\caption{\textit{Schematic Representation of Institutionally Mediated Private
Value Exchange: Nodes.}  The distributed ledger is operated by a federation of
regulated institutions.}

\label{f:peven}
\end{center}
\end{figure}

Our second approach starts with the assumption that the ``public''
cryptocurrencies are not suitable for all kinds of institutional support,
perhaps for the reasons cited in Section~\ref{ss:isc}.  Instead, it proposes to
establish a distributed ledger for conducting financial transactions, and that
each node of the distributed ledger would be owned and operated by a regulated
institution, as shown in Figure~\ref{f:peven}.  This could be achieved with a
``permissioned'' distributed ledger system such as
Hyperledger~\cite{hyperledger}, using an energy-efficient Byzantine
fault-tolerant consensus algorithm such as PBFT~\cite{castro1999}.  Users and
governments would benefit from the fact that transacting parties would not need
to use cryptocurrency of dubious value but in fact could transact using digital
versions of state-issued currency, i.e. \textit{central bank digital currency}
(CBDC), which is currently under consideration by central banks around the
world and may offer a variety of economic and operational
benefits~\cite{bis2018}.

At this point it might be tempting to suggest that since the entire network
consists of regulated or otherwise approved financial institutions, then
governments should require the establishment of a ``master key'' or other
exceptional access mechanism, so that they might be able to break the anonymity
of users.  We argue that this temptation should be resisted.  Over the years,
policymakers have called for broadly applied exceptional access mechanisms in a
variety of contexts, and after considerable debate, such calls have been found
to be premature and subsequently
withdrawn~\cite{abelson1997,abelson2015,benaloh2018}.  Indeed, legislators in
the United States~\cite{hr5823} and France~\cite{thomson2016} have gathered
opposition to exceptional access mechanisms, citing their intrinsic security
weaknesses and potential for abuse.

Indeed, for the approach we present to be a private value exchange, the
regulated institutions must \textit{commit to facilitating private
transactions}.  At one level, the institutions must adopt the specific
technologies such as ring signatures, stealth addresses, and confidential
transactions used by privacy-enabling cryptocurrencies such as Monero.  At
another level, the institutions must commit to an ongoing effort to audit,
challenge, and improve the technology and operational procedures, because
privacy-enhancing technologies require vigilance~\cite{zimmermann1991}.  It
follows that the institutions and the authorities of the jurisdictions in which
they operate must commit to ensuring that the technology and operational
procedures are effective in safeguarding the privacy of transacting parties
against politically, financially, and technologically powerful groups who might
have contrary interests.

It is assumed that authorities would take the same measures described in
Section~\ref{ss:isc} to ensure that corporations and registered businesses use
known, monitorable accounts for all of their transactions.  Enforcement of such
a policy would be qualitatively easier in this case since the entire network is
owned and operated by regulated institutions, and regulators could expect the
same benefits associated with monitoring taxable income and reconciling line
items in cash flow statements against actual, auditable transfers on the
distributed ledger.

\begin{figure}
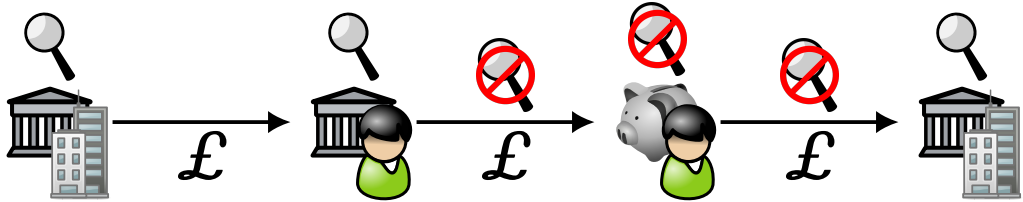

\begin{center}
\begin{tikzpicture}[>=latex, node distance=3cm, font={\sf \small}, auto]\ts
\node (r1) at (0,0) [noshape, text width=4em] {
    \scalebox{0.8}{\includegraphics{images/primary_bank.pdf}}
};
\node (o1) at (0.4,-0.3) [noshape, text width=4em] {
    \scalebox{0.9}{\includegraphics{images/office-towers.pdf}}
};
\node (r2) at (4,0) [noshape, text width=4em] {
    \scalebox{0.8}{\includegraphics{images/primary_bank.pdf}}
};
\node (o2) at (4.4,-0.4) [noshape, text width=4em] {
    \scalebox{0.4}{\includegraphics{images/people-juliane-krug-08a.pdf}}
};
\node (r3) at (8,0) [noshape, text width=4em] {
    \scalebox{1.2}{\includegraphics{images/gray_piggybank.pdf}}
};
\node (o3) at (8.4,-0.4) [noshape, text width=4em] {
    \scalebox{0.4}{\includegraphics{images/people-juliane-krug-08a.pdf}}
};
\node (r4) at (12,0) [noshape, text width=4em] {
    \scalebox{0.8}{\includegraphics{images/primary_bank.pdf}}
};
\node (o4) at (12.4,-0.3) [noshape, text width=4em] {
    \scalebox{0.9}{\includegraphics{images/office-towers.pdf}}
};
\node (m1) at (0,1) [noshape] {
    \scalebox{0.06}{\includegraphics{images/office-glass-magnify.pdf}}
};
\node (m2) at (4,1) [noshape] {
    \scalebox{0.06}{\includegraphics{images/office-glass-magnify.pdf}}
};
\node (m2) at (12,1) [noshape] {
    \scalebox{0.06}{\includegraphics{images/office-glass-magnify.pdf}}
};
\node (m1) at (8,1.1) [noshape] {
    \scalebox{0.06}{\includegraphics{images/office-glass-magnify-no.png}}
};

\draw[->, line width=0.5mm] (r1) -- node[below] {
    \textbf{\textit{\Huge \pounds}}
} node[above] {
} (r2);
\draw[->, line width=0.5mm] (r2) -- node[below] {
    \textbf{\textit{\Huge \pounds}}
} node[above] {
    \scalebox{0.06}{\includegraphics{images/office-glass-magnify-no.png}}
} (r3);
\draw[->, line width=0.5mm] (r3) -- node[below] {
    \textbf{\textit{\Huge \pounds}}
} node[above] {
    \scalebox{0.06}{\includegraphics{images/office-glass-magnify-no.png}}
} (r4);

\end{tikzpicture}

\caption{\textit{Schematic Representation of Institutionally Mediated Private
Value Exchange: Private Transactions.}  As in Figure~\ref{f:atf1}, an
individual receives funds into her institutional account (second icon from
left) and transfers them to her private store (second icon from right).  Unlike
in Figure~\ref{f:atf1}, the funds may be state-issued currency, as indicated by
the Pound Sterling symbols, rather than cryptocurrency.  When she wants to make
a payment, she must remit it from her private store to an account held by a
regulated institution (rightmost icon).}

\label{f:pevept}
\end{center}
\end{figure}

State actors would realise another important benefit from this approach as
well.  Because all transactions must necessarily involve a regulated
institution, transactions of the sort described in Figure~\ref{f:atf3}, in
which private actors exchange value directly via their own private stores,
would not be possible.  Figure~\ref{f:pevept} illustrates how a user would make
payments privately.  A user would initially receive funds into her account with
a registered institution, which she would in turn remit to her private store.
When she wants to make a payment to a merchant or service provider, she can
remit the funds to the account that that organisation holds with a registered
institution.  The privacy features of the distributed ledger, such as ring
signatures, stealth addresses, transaction confidentiality, and any other
necessary features that may be developed from time to time, would ensure that
when the individual makes the payment, she does not reveal either her identity
or any information about her prior transactions, including the transactions
from which she originally received the funds.

\begin{figure}
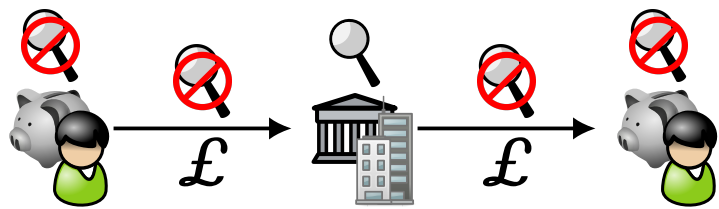

\begin{center}
\begin{tikzpicture}[>=latex, node distance=3cm, font={\sf \small}, auto]\ts
\node (r1) at (0,0) [noshape, text width=4em] {
    \scalebox{1.2}{\includegraphics{images/gray_piggybank.pdf}}
};
\node (o1) at (0.4,-0.4) [noshape, text width=4em] {
    \scalebox{0.4}{\includegraphics{images/people-juliane-krug-08a.pdf}}
};
\node (r2) at (4,0) [noshape, text width=4em] {
    \scalebox{0.8}{\includegraphics{images/primary_bank.pdf}}
};
\node (o2) at (4.4,-0.3) [noshape, text width=4em] {
    \scalebox{0.9}{\includegraphics{images/office-towers.pdf}}
};
\node (r3) at (8,0) [noshape, text width=4em] {
    \scalebox{1.2}{\includegraphics{images/gray_piggybank.pdf}}
};
\node (o3) at (8.4,-0.4) [noshape, text width=4em] {
    \scalebox{0.4}{\includegraphics{images/people-juliane-krug-08a.pdf}}
};
\node (m1) at (0,1.1) [noshape] {
    \scalebox{0.06}{\includegraphics{images/office-glass-magnify-no.png}}
};
\node (m2) at (4,1) [noshape] {
    \scalebox{0.06}{\includegraphics{images/office-glass-magnify.pdf}}
};
\node (m3) at (8,1.1) [noshape] {
    \scalebox{0.06}{\includegraphics{images/office-glass-magnify-no.png}}
};

\draw[->, line width=0.5mm] (r1) -- node[below] {
    \textbf{\textit{\Huge \pounds}}
} node[above] {
    \scalebox{0.06}{\includegraphics{images/office-glass-magnify-no.png}}
} (r2);
\draw[->, line width=0.5mm] (r2) -- node[below] {
    \textbf{\textit{\Huge \pounds}}
} node[above] {
    \scalebox{0.06}{\includegraphics{images/office-glass-magnify-no.png}}
} (r3);

\end{tikzpicture}

\caption{\textit{Schematic Representation of Institutionally Mediated Private
Value Exchange: Mediated Transactions Between Consumers.}  Individuals (outer
icons) wishing to transact with each other via their private stores rather than
accounts with regulated institutions must transact via a regulated intermediary
(centre icon).}

\label{f:pevmed}
\end{center}
\end{figure}

By ensuring that no single enterprise receives too large a share of any
individual's transactions in the system, the use of a distributed ledger
achieves an essential requirement of the design.  Individuals would be expected
to use their private stores to transact with many different counterparties, via
their own regulated intermediaries, so no single intermediary would have a
global, ``panopticon-like'' view of all of the individual's transactions.

Since individuals cannot transact directly via their private stores, to
exchange value they must transact via a regulated intermediary as shown in
Figure~\ref{f:pevmed}.  Individuals conducting transactions might not need to
have accounts to exchange value with each other; we surmise that the regulated
intermediary would perform the service for a fee.  We also suggest that the
intermediary would not be required to carry out strong identification of the
sort required by the FATF recommendations~\cite{fatf-recommendations} but might
require a less-stringent form of identification, such as an attribute-backed
credential indicating that either the sender or the receiver are eligible to
transact~\cite{camenisch2013}.  Regulated intermediaries could also provide
token mixing services for groups of individuals who satisfy AML criteria,
without explicitly requiring knowledge of their unitary identities.

If successfully operationalised, the approach described in this section would
offer governments the same benefits to taxation and auditing as the approach
described in Section~\ref{ss:isc}, and governments would additionally gain the
ability to impose blacklists or economic sanctions on targeted recipients.
Individuals would receive the same privacy benefits described in
Section~\ref{ss:isc} for transactions involving merchants and service
providers, and identification requirements of intermediaries for other
transactions could be made parsimonious.  However, there are two main drawbacks
for individuals seeking privacy, the first being that individuals would need to
interact with a registered intermediary before they are able to make or receive
payments.  The other, more serious concern is the question of the mechanism by
which the privacy-enabling properties of the system is assured.  Inasmuch as
cryptocurrencies represent a check on state power~\cite{sotirakopoulos2017}, we
have reason to believe that the privacy characteristics of cryptocurrencies
will continue to improve, despite their demonstrable
shortcomings~\cite{moser2018,kappos2018}.

If the regulated institutions that design, deploy, and maintain the
infrastructure for executing transactions are asked to carry the flag for the
privacy of their clients, then there could be a misalignment of interests.
Clients would need to know the actual privacy limitations of the
infrastructure, so adversarial audits would need to be carried out from time to
time in the interest of the public.  Then, institutions would need incentives
and resources to continuously improve the infrastructure and fix any
deficiencies on an ongoing basis.  A process for admitting new participants
would be necessary to ensure that the network remains distributed, and it would
need to satisfy an openness criterion to ensure that privacy-threatening
procedures do not develop outside the view of the public eye.  There would also
need to be a diversity of implementations, such that sporadic vulnerabilities
do not threaten the privacy of a significant share of the users of the system.

Arguably, such incentives exist among cryptocurrencies, since they must compete
for business.  It remains to be seen whether effective auditing and competition
could assure the privacy-enabling properties of a value exchange operated
entirely by institutions.


\section{Conclusions}

\begin{table}
\begin{center}
\sf\begin{tabular}{|L{4cm}|p{1.1cm}p{1.1cm}p{1.2cm}p{1.2cm}p{1.2cm}|p{1.2cm}p{1.2cm}|}\hline
& \rotatebox{90}{cash}
& \rotatebox{90}{modern} \rotatebox{90}{retail banking}
& \rotatebox{90}{``traditional''} \rotatebox{90}{cryptocurrency} \rotatebox{90}{(e.g. Bitcoin)}
& \rotatebox{90}{``traditional''} \rotatebox{90}{stablecoins} \rotatebox{90}{(e.g. Tether)}
& \rotatebox{90}{privacy-enabling} \rotatebox{90}{cryptocurrency} \rotatebox{90}{(e.g. Monero)}
& \rotatebox{90}{inst. supported\hspace{8pt}} \rotatebox{90}{privacy-enabling} \rotatebox{90}{cryptocurrency}
& \rotatebox{90}{institutionally} \rotatebox{90}{mediated private\hspace{8pt}} \rotatebox{90}{value exchange}\\
Robust to cyberattacks      & \CIRCLE & \Circle & \Circle & \Circle & \Circle & \Circle & \Circle \\
Usable without registration & \CIRCLE & \Circle & \CIRCLE & \CIRCLE & \CIRCLE & \CIRCLE & \Circle \\
Unlinkable* transactions    & \RIGHTcircle & \Circle & \Circle & \Circle & \CIRCLE & \CIRCLE & \CIRCLE \\
Electronic transactions     & \Circle & \CIRCLE & \CIRCLE & \CIRCLE & \CIRCLE & \CIRCLE & \CIRCLE \\
Suitable for taxation       & \RIGHTcircle & \CIRCLE & \Circle & \RIGHTcircle & \Circle & \CIRCLE & \CIRCLE \\
Can block some illicit uses & \Circle & \CIRCLE & \Circle & \Circle & \Circle & \Circle & \CIRCLE \\
Can be denominated in       & \CIRCLE & \CIRCLE & \Circle & \CIRCLE & \Circle & \Circle & \CIRCLE \\
\vspace{-4.5mm}\hspace{3mm} units of fiat currency & & & & & & & \\
\hline\end{tabular}
\vspace{-4pt}
\end{center}

\sf\small\hspace{5pt}*Potentially\rm

\caption{\textit{Comparison of various electronic payment methods, including
the new proposed methods.}}

\label{t:peeve}
\end{table}

Framing the ongoing conversation about the future of payments as a set of
tradeoffs, we introduced two possible candidate architectures for a
privacy-enabling electronic value exchange: \textit{institutionally supported
privacy-enabling cryptocurrency} and \textit{institutionally mediated private
value exchange}.  Both architectures require both the design, implementation,
deployment, and maintenance of new technology as well as the development of
regulatory policy in which such technology will operate.  Table~\ref{t:peeve}
summarises the tradeoffs and contextualises our two prospective approaches.
Cash has many desirable properties, such as universality (i.e., its use does
not require a relationship with a registered institution) and privacy in
practice (serial numbers on banknotes can be traced but generally are not).
However, it cannot be sent across computer networks and is sometimes used for
illicit transactions, including tax evasion.  In contrast, modern retail
banking requires accounts and facilitates large-scale surveillance.  The most
popular cryptocurrencies such as Bitcoin do not actually avoid surveillance and
are in some ways potentially easier to trace than ordinary retail transactions.
Privacy-enabling cryptocurrencies promise to address both deficiencies,
although research has shown that the goals motivating their development have
not yet been fully achieved.

The various approaches to electronic payments each have their own advantages
and limitations, and by elaborating the tradeoffs, we hope to facilitate a more
fulsome conversation among the stakeholders and offer a useful framework for
discussing future solutions.  We believe that both approaches have their place
and prospective adherents, and the adoption of one would not exclude the
adoption of the other.  Businesses that offer services to cryptocurrency users
and traders would find value in the first approach, and businesses seeking to
facilitate private, cash-like electronic transactions within a regulated system
would find value in the second approach.  Correspondingly, some regulators
might be troubled by supporting trade in assets whose value and uses are beyond
their reach, as would be the case in the first approach, and some
privacy-minded individuals might be troubled by the possibility that the
regulated financial institutions that operate the system described in the
second approach might secretly collude to compromise the anonymity of their
clients.

We suggest that institutionally supported privacy-enabling cryptocurrency would
be strictly better than privacy-enabling cryptocurrency without institutional
support, mainly because regulators would benefit from the ability to monitor
corporations and registered businesses that use cryptocurrencies.  We also
suggest that institutionally mediated private value exchange would be strictly
better than modern retail banking as currently practiced, mainly because users
would avoid payment networks and enjoy an improved expectation of privacy in
their ordinary activities.  However, neither approach achieves all of the
objectives of both parties.  For example, the ability to transact without
interacting with a regulated institution may be incompatible with the ability
for a government to block illicit use.  Similarly, monetary policy might not be
possible if cryptocurrency governance were exogenous to the state, although the
possibility of this happening at scale seems remote.  As the hard choices for
the future of payments come to light, we believe that acknowledgment and
discussion of these tradeoffs, as well as a commitment to both serious privacy
and serious regulation, are prerequisites for advancing the interests of all
stakeholders.

\section*{Acknowledgements}

The authors would like to thank Edgar Whitley and David Pym for their
insightful contributions.  Geoff Goodell is also an associate of the Centre for
Technology and Global Affairs of the University of Oxford.  We acknowledge the
Engineering and Physical Sciences Research Council (EPSRC) for the BARAC
project (EP/P031730/1) and the European Commission for the FinTech project
(H2020-ICT-2018-2 825215). Tomaso Aste acknowledges the Economic and Social
Research Council (ESRC) for funding the Systemic Risk Centre (ES/K0 02309/1).

\footnotesize\noindent Diagram Clip Art Image Credits: \texttt{publicdomainvectors.org, bitcoin.org, twitter.com}

\end{document}